\documentclass[times, 10pt, twocolumn, 3p]{elsarticle}

\usepackage{times}
\usepackage{graphicx}
\usepackage{amssymb}
\usepackage{gensymb}
\usepackage{amsmath}

\usepackage{afterpage}

\usepackage{url}
\usepackage{hyperref} 
\usepackage{breakurl}

\hypersetup{
colorlinks,
linkcolor=blue,
citecolor=blue,
filecolor=blue,
urlcolor=blue}
\usepackage[labelfont={bf},font=small]{caption}
\usepackage[none]{hyphenat}

\usepackage{mathtools, cuted}

\usepackage{multirow}

\usepackage[usenames, dvipsnames]{color}

\usepackage{float}

\usepackage{pifont}

\usepackage{tikz}

\usepackage[framemethod=tikz]{mdframed}
            
\newcommand*\diff{\mathop{}\!\mathrm{d}}

\begin{document}

\begin{frontmatter}

\title{Exit Wavefunction Reconstruction from Single Transmission Electron Micrographs with Deep Learning}

\author[a1]{Jeffrey M. Ede\corref{cor1}}
\ead{j.m.ede@warwick.ac.uk}

\author[a1]{Jonathan J. P. Peters}
\ead{j.peters.1@warwick.ac.uk}

\author[a1]{Jeremy Sloan}
\ead{j.sloan@warwick.ac.uk}

\author[a1]{Richard Beanland}
\ead{r.beanland@warwick.ac.uk}

\cortext[cor1]{Corresponding author}
\address[a1]{Department of Physics, University of Warwick, Coventry, England, CV4 7AL}

\begin{abstract}

Half of wavefunction information is undetected by conventional transmission electron microscopy (CTEM) as only the intensity, and not the phase, of an image is recorded. Following successful applications of deep learning to optical hologram phase recovery, we have developed neural networks to recover phases from CTEM intensities for new datasets containing 98340 exit wavefunctions. Wavefunctions were simulated with clTEM multislice propagation for 12789 materials from the Crystallography Open Database. Our networks can recover 224$\times$224 wavefunctions in $\sim$25 ms for a large range of physical hyperparameters and materials, and we demonstrate that performance improves as the distribution of wavefunctions is restricted. Phase recovery with deep learning overcomes the limitations of traditional methods: it is live, not susceptible to distortions, does not require microscope modification or multiple images, and can be applied to any imaging regime. This paper introduces multiple approaches to CTEM phase recovery with deep learning, and is intended to establish starting points to be improved upon by future research. Source code and links to our new datasets and pre-trained models are available at \url{https://github.com/Jeffrey-Ede/one-shot}.

\end{abstract}


\end{frontmatter}

\section{Introduction}

\begin{figure}[tbh!]
\centering
\includegraphics[width=\columnwidth]{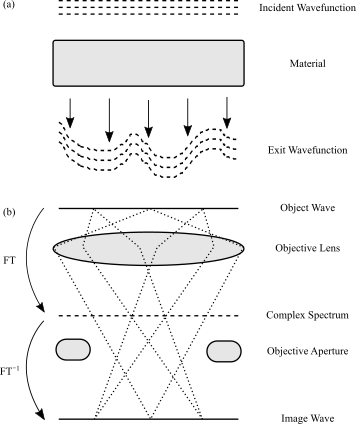}
\caption{ Wavefunction propagation. a) An incident wavefunction is perturbed by a projected potential of a material. b) Fourier transforms (FTs) can describe a wavefunction being focused by an objective lens through an objective aperture to a focal plane. }
\label{fig:microscope}
\end{figure}

Information transfer by electron microscope lenses and correctors can be described by wave optics\cite{lehmann2002tutorial} as electrons exhibit wave-particle duality\cite{frabboni2007young, matteucci1998experiment}. In a model electron microscope, a system of condenser lenses directs electrons illuminating a material into a planar wavefunction, $\psi_\text{inc}(\textbf{r}, z)$, with wavevector, $\textbf{k}$. Here, $z$ is distance along its optical axis in the electron propagation direction, described by unit vector $\hat{\textbf{z}}$, and $\textbf{r}$ is the position in a plane perpendicular to the optical axis. As $\psi_\text{inc}(\textbf{r}, z)$ travels through a material in fig.~\ref{fig:microscope}a, it is perturbed to an exit wavefunction, $\psi_\text{exit}(\textbf{r}, z)$, by a material potential.

The projected potential of a material in direction $\hat{\textbf{z}}$, $U(\textbf{r}, z)$, and corresponding structural information can be calculated from $\psi_\text{exit}(\textbf{r}, z)$\cite{lentzen2000reconstruction, auslender2019measuring}. For example, 
\begin{equation}
U(\textbf{r}) \approx \dfrac{\text{Im}(\psi_\text{exit}(\textbf{r}, z) \exp( i \varphi) - \langle \psi_\text{exit}(\textbf{r}, z)\rangle_\textbf{r})}{\lambda\xi\sin(\pi z/\xi)},
\end{equation}
for a typical crystal system well-approximated by two Bloch waves\cite{lentzen2000reconstruction}. Here $\varphi$ is a distance between Bloch wavevectors, $\lambda$ is the electron wavelength, $\xi$ is an extinction distance for two Bloch waves, $\langle...\rangle_\textbf{r}$ denotes an average with respect to $\textbf{r}$, and $\text{Im}(z)$ is the imaginary part of $z$. Other applications of $\psi_\text{exit}(\textbf{r}, z)$\cite{tonomura1987applications} include information storage, point spread function deconvolution, improving contrast, aberration correction\cite{fu1991correction}, thickness measurement\cite{mccartney1994absolute}, and electric and magnetic structure determination\cite{park2014observation, trippoff}. Exit wavefunctions can also simplify comparison with simulations as no information is lost.

In general, the intensity, $I(S)$, of a measurement with support, $S$, is
\begin{equation}\label{eqn:intensity_integral}
I(S) = \int\limits_{s\in S} |\psi(s)|^2 \diff{s}.
\end{equation}
A support is a measurement region, such as an electron microscope camera\cite{mcmullan2016direct, mcmullan2009detective} element. Half of wavefunction information is lost at measurement as $|\psi|^2$ is a function of amplitude, $A > 0$, and not phase, $\theta \in [-\pi, \pi)$, \begin{equation}
|\psi|^2 = |A\exp( i \theta)|^2 = A^2|\exp( i \theta)|^2 = A^2.
\end{equation}
We emphasize that we define $A$ to be positive so that $|\psi|^2 \mapsto A$ is bijective, and $\psi$ sign information is in $\exp( i \theta)$. Phase information loss is a limitation of conventional single image approaches to electron microscopy, including transmission electron microscopy\cite{carter2016transmission} (TEM), scanning transmission electron microscopy\cite{pennycook2011scanning} (STEM), and scanning electron microscopy\cite{goldstein2017scanning} (SEM).


In the Abbe theory of wave optics\cite{kohler1981abbe} in fig.~\ref{fig:microscope}b, the projection of $\psi$ to a complex spectrum, $\psi_\text{dif}(\textbf{q})$, in reciprocal space, $\textbf{q}$, at the back focal plane of an objective lens can be described by a Fourier transform (FT) 
\begin{equation} 
\psi_\text{dif}(\textbf{q}) = \text{FT}[\psi_\text{exit}(\textbf{r})] = \int\limits \psi_\text{exit}(\textbf{r})\exp(-2\pi i \textbf{q} \cdot \textbf{r}) \diff{\textbf{r}}.
\end{equation} 
In practice, $\psi_\text{dif}(\textbf{q})$ is perturbed to $\psi_\text{pert}$ by an objective aperture, $E_\text{ap}$, coherence, $E_\text{coh}$, chromatic aberration, $E_\text{chr}$, and lens aberrations, $\chi$, and is described in the Fourier domain\cite{lehmann2002tutorial} by
\begin{align}
\psi_\text{pert}(\textbf{q}) &= E_\text{ap}(\textbf{q}) E_\text{coh}(\textbf{q}) E_\text{chr}(\textbf{q})  
\exp(- i \chi(\textbf{q})) \psi_\text{dif}(\textbf{q}) 
\end{align}
where 
\begin{align}
E_\text{ap}(\textbf{q}) &= \begin{cases}
    1, & \text{for } |\textbf{q}| \le k\theta_\text{max} \\
    0, & \text{for } |\textbf{q}| > k\theta_\text{max} \\
  \end{cases} \\
E_\text{coh}(\textbf{q}) &= \exp\left(-\dfrac{(\nabla\chi(\textbf{q}))^2(k\theta_\text{coh})^2}{4\ln(2)}\right) \\
E_\text{chr}(\textbf{q}) &= \exp\bigg( -\dfrac{1}{2}\bigg( \pi k C_\text{c} \dfrac{\Delta E}{U_a^*} \bigg( \dfrac{q}{k} \bigg)^2 \bigg)^2 \bigg)
\end{align}
\begin{align}
\begin{split}
\chi(\theta, \phi) = \sum\limits_{n=0}^\infty \sum\limits_{m=0}^{n+1} \ &\dfrac{C_{n,m,a}\theta^{n+1}\cos(m\phi)}{n+1} \ + \\ 
& \dfrac{C_{n,m,b}\theta^{n+1}\sin(m\phi)}{n+1}
\end{split}
\end{align}
for an objective aperture with angular extent, $\theta_\text{max}$, illumination aperture with angular extent, $\theta_\text{coh}$, energy spread, $\Delta E$, chromatic aberration coefficient of the objective lens, $C_\text{c}$, relativistically corrected acceleration voltage, $U_a^*$, aberration coefficients, $C_{n,m,a}$ and $C_{n,m,b}$, angular inclination of perturbed wavefronts to the optical axis, $\phi$, angular position in a plane perpendicular to the optical axis, $\theta$, $m, n \in \mathbb{N}_0$, and $m+n$ is odd.


All waves emanating from points in Fourier space interfere in the image plane to produce an image wave, $\psi_\text{img}(\textbf{r})$, mathematically described by an inverse Fourier transform (FT$^{-1}$) 
\begin{equation}
\psi_\text{img}(\textbf{r}) = \text{FT}^{-1}(\psi_\text{pert}(\textbf{q})) = \int \psi_\text{pert}(\textbf{q}) \exp(2\pi i \textbf{q} \cdot \textbf{r}) \diff{\textbf{q}}.
\end{equation}

Information transfer from $\psi_\text{exit}$ to measured intensities can be modified by changing $\chi$. Typically, by controlling the focus of the objective lens. However, half of $\psi_\text{exit}$ information is missing from each measurement. To overcome this limitation, a wavefunction can be iteratively fitted to a series of aligned images with different $\chi$\cite{lubk2016fundamentals, koch2010off, koch2014towards, haigh2013recording}. However, collecting an image series, waiting for sample drift to decay, and iterative fitting delays each $\psi_\text{exit}$ measurement. As a result, aberration series reconstruction is unsuitable for live exit wavefunction reconstruction.

Electron holography\cite{lehmann2002tutorial, koch2010off, ozsoy2014hybridization} is an alternative approach to exit wavefunction reconstruction that compares $\psi_\text{exit}$ to a reference wave. Typically, a hologram, $I_\text{hol}$, is created by moving a material off-axis and introducing an electrostatic biprism after the objective aperture. The Fourier transform of a M\"ollenstedt biprismatic hologram is\cite{lehmann2002tutorial}
\begin{align}\label{eqn:hologram}
\begin{split}
\text{FT}(I_\text{hol}(\textbf{r})) = \ & \text{FT}(1+|\psi_\text{exit}(\textbf{r})|^2) \ + \\
 & \mu \text{FT}( \psi_\text{exit}(\textbf{r}) ) \otimes \delta(\textbf{q}-\textbf{q}_c) \ + \\
 & \mu \text{FT}( \psi_\text{exit}^*(\textbf{r}) ) \otimes \delta(\textbf{q}+\textbf{q}_c),
\end{split}
\end{align}
where $\psi_\text{exit}^*(\textbf{r})$ is the complex conjugate of $\psi_\text{exit}(\textbf{r})$, $|\textbf{q}_c|$ is the carrier frequency of interference fringes, and their contrast,
\begin{equation}
\mu = |\mu_\text{coh}||\mu_\text{inel}||\mu_\text{inst}| \textit{MTF},
\end{equation}
is given by source spatiotemporal coherence, $\mu_\text{coh}$, inelastic interactions, $\mu_\text{inst}$, instabilities, $\mu_\text{inst}$, and the modulation transfer function\cite{ruskin2013quantitative}, $\textit{MTF}$, of a detector. Convolutions with Dirac $\delta$ in eqn.~\ref{eqn:hologram} describe sidebands in Fourier space that can be cropped, centered, and inverse Fourier transformed for live exit wavefunction reconstruction. However, off-axis holograms are susceptible to distortions and require meticulous microscope alignment as phase information is encoded in interference fringes\cite{lehmann2002tutorial}, and cropping Fourier space reduces resolution\cite{ozsoy2014hybridization}. 

Artificial neural networks (ANNs) have been trained to recover phases of optical holograms from single images\cite{rivenson2018phase}. In general, this is not possible as there are an infinite number of physically possible $\theta$ for a given $A$. However, ANNs are able to leverage an understanding of the physical world to recover $\theta$ if the distribution of possible holograms is restricted, for example, to biological cells.  Non-iterative methods that do not use ANNs to recover phase information from single images have also been developed. However, they are limited to defocused images in the Fresnel regime\cite{morgan2011direct}, or to non-planar incident wavefunctions in the Fraunhofer regime\cite{martin2008direct}.

One-shot phase recovery with ANNs overcomes the limitations of traditional methods: it is live, not susceptible to off-axis holographic distortions, does not require microscope modification, and can be applied to any imaging regime. In addition, ANNs could be applied to recover phases of images in large databases, long after samples may have been lost or destroyed. In this paper, we investigate the application of deep learning to one-shot exit wavefunction reconstruction in conventional transmission electron microscopy (CTEM).

\section{Exit Wavefunction Datasets}

\begin{figure*}[tbh!]
\footnotesize
\begin{tabular*}{\textwidth}{l@{\extracolsep{\fill}}cccccc}
\hline
\multicolumn{1}{c}{Dataset} & $n$ & Train  & Unseen & Validation & Test & Total \\ 
\hline
Multiple Materials & 1 & 25325 & 1501 & 3569 & 8563 & 38958 \\
Multiple Materials & 3 & 24530 & 1544 & 3399 & 8395 & 37868 \\
Multiple Materials, Restricted & 3 & 8002 & - & 1105 & 2763 & 11870 \\
In$_{1.7}$K$_2$Se$_8$Sn$_{2.28}$ & 1 & 3856 & - & 963 & - & 4819 \\
In$_{1.7}$K$_2$Se$_8$Sn$_{2.28}$ & 3 & 3861 & - & 964 & - & 4825 \\
\hline
\end{tabular*}
\captionof{table}{ New datasets containing 98340 wavefunctions simulated with clTEM are split into training, unseen, validation, and test sets. Unseen wavefunctions are simulated for training set materials with different simulation hyperparameters. Kirkland potential summations were calculated with $n=3$ or truncated to $n=1$ terms, and dashes (-) indicate subsets that have not been simulated. Datasets have been made publicly available at \cite{warwickem!}. }
\label{table:new_datasets}
\end{figure*}

To showcase one-shot exit wavefunction reconstruction, we generated 98340 exit wavefunctions with clTEM\cite{clTEM_repo, dyson2014advances} multislice propagation for 12789 CIFs\cite{hall1991crystallographic} downloaded from the Crystallography Open Database\cite{Quiros2018, Merkys2016, Grazulis2015, Grazulis2012, Grazulis2009, Downs2003} (COD). Complex 64 bit 512$\times$512 wavefunctions were simulated for CTEM with acceleration voltages in $\{80, 200, 300\}$ kV, material depths along the optical axis uniformly distributed in $[5, 100)$ nm, material widths perpendicular to the optical axis in $[5, 10)$ nm, and crystallographic zone axes $(h, k, l)$ $h, k, l \in \{0, 1, 2\}$. Materials are padded on all sides with 0.8 nm of vacuum in the image plane, and 0.3 nm along the optical axis, to reduce simulation artefacts. Finally, crystal tilts to each axis were perturbed by zero-centered Gaussian random variates with standard deviation 0.1$\degree$. We used default values for other clTEM hyperparameters. 


Multislice exit wavefunction simulations with clTEM are based on \cite{kirkland2010advanced}. Simulations start with a planar wavefunction, $\psi$, travelling along a TEM column
\begin{equation}
\psi \left( x, y, z \right) = \exp \left( \frac{2 \pi i z}{\lambda} \right),
\label{eq:planewave}
\end{equation}
where $x$ and $y$ are in-plane coordinates, and $z$ is distance travelled. After passing through a thin specimen, with thickness $\Delta z$, wavefunctions are approximated by 
\begin{equation}
\psi \left( x, y, z + \Delta z \right) \simeq \exp \left( i \sigma V_z\left( x, y \right) \Delta z \right) \psi \left( x, y, z \right)
\label{eq:weak_phase}
\end{equation}
with
\begin{equation}
\sigma = \frac{2 \pi m e \lambda}{h^2},
\end{equation}
where $V_z$ is the projected potential of the specimen at $z$, $m$ is relativistic electron mass, $e$ is fundamental electron charge, and $h$ is Planck's constant. 

For electrons propagating through a thicker specimen, cumulative phase change can described by a specimen transmission function, $t(x,y,z)$, so that
\begin{equation}
\psi \left( x, y, z + \Delta z \right) =  t\left( x, y, z \right) \psi \left( x, y, z \right)
\end{equation}  
with
\begin{equation}
t \left( x, y, z \right) = \exp \left( i \sigma \int \limits_z^{z+\Delta z} V\left( x, y, z' \right) \diff{z'}\right).
\end{equation}

A thin sample can be divided into multiple thin slices stacked together using a propagator function, $P$, to map wavefunctions between slices. A wavefunction at slice $n$ is mapped to a wavefunction at slice $n+1$ by
\begin{equation}
\psi_{n+1} \left( x, y \right) \leftarrow P \left( x, y, \Delta z \right) \otimes \left[ t_n \left( x, y \right) \psi_n \left( x, y, \right) \right]
\end{equation}
where $\psi_0$ is the incident wave in eqn.~\ref{eq:planewave}. Simulations with clTEM are based on OpenCL\cite{stone2010opencl}, and use  graphical processing units (GPUs) to accelerate fast Fourier transform\cite{moreland2003fft} (FFT) based convolutions. The propagator is calculated in reciprocal space
\begin{equation}
P \left( k_x, k_y \right) = \exp \left( -i \pi \lambda k^2 \Delta z \right),
\end{equation}
where $k_x$, $k_y$ are reciprocal space coordinates, and $k = ( k_x^2 + k_y^2 )^{1/2}$. As Fourier transforms are used to map between reciprocal and real space, propagator and transmission functions are band limited to decrease aliasing.

Projected atomic potentials are calculated using Kirkland's parameterization\cite{kirkland2010advanced}, where the projected potential of an atom at position, $p$, in a thin slice is
\begin{align}\label{eqn:kirkland_sum}
\begin{split}
v_p \left( x, y \right) = \, & 4 \pi^2 e r_\text{Bohr} \sum_i^n a_i K_0 \left( 2 \pi r_p b_i^{1/2} \right) \ + \\
& 2 \pi^2 e r_\text{Bohr} \sum_i^n \frac{c_i}{d_i} \exp\left( - \dfrac{\pi^2 r_p^2}{d_i} \right), 
\end{split}
\end{align}
where $r_p = [(x-x_p)^2 + (y-y_p)^2]^{1/2}$, $x_p$ and $y_p$ are the coordinates of the atom, $r_\text{Bohr}$ is the Bohr radius, $K_0$ is the modified Bessel function\cite{abramowitz1964handbook}, and the parameters $a_i$, $b_i$, $c_i$, and $d_i$ are tabulated for each atom in \cite{kirkland2010advanced}. Nominally, $n=3$. However, we also use $n=1$ to investigate robustness to alternative simulation physics. In effect, simulations with $n=1$ are for an alternative universe where atoms have different potentials. Every atom in a slice contributes to the total projected potential
\begin{equation}
V_z = \sum_p v_p.    
\end{equation}

After simulation, a 320$\times$320 region was selected from the center of each wavefunction to remove edge artefacts. Each wavefunction was divided by its magnitude to prevent an ANN from inferring information from an absolute intensity scale. In practice, it is possible to measure an absolute scale; however, it is specific to a microscope and its configuration. 

To investigate ANN performance for multiple materials, we partitioned 12789 CIFs into training, validation, and test sets by journal of publication. There are 8639 training set CIFs: 150 New Journal of Chemistry, 1034 American Mineralogist, 1998 Journal of the American Chemical Society, and 5457 Inorganic Chemistry. In addition, there are 1216 validation set CIFs published in Physics and Chemistry of Materials, and 2927 test set CIFs published in Chemistry of Materials. Wavefunctions were simulated for three random sets of hyperparameters for each CIF, except for a small portion of examples that were discarded because CIF format or simulation hyperparameters were unsupported. Partitioning by journal helps to test the ability of an ANN to generalize given that wavefunction characteristics are expected to vary with journal. 

\begin{figure}[tbp!]
\centering
\includegraphics[width=0.95\columnwidth]{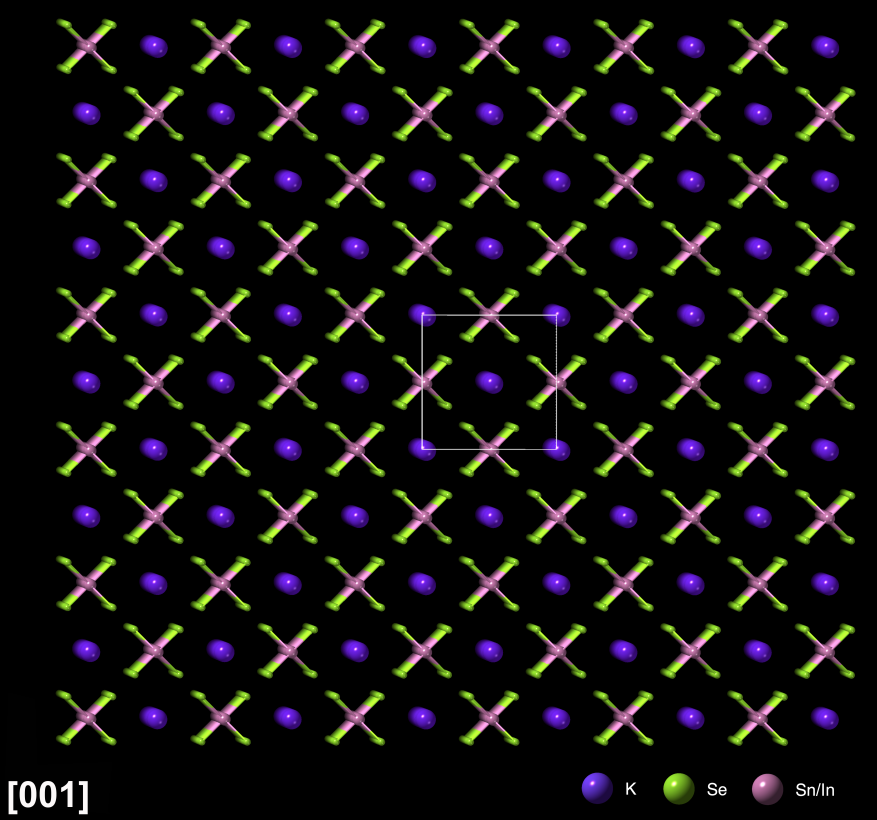}
\caption{ Crystal structure of In$_{1.7}$K$_2$Se$_8$Sn$_{2.28}$ projected along Miller zone axis [001]. A square outlines a unit cell. }
\label{fig:random_material}
\end{figure}

New simulated wavefunction datasets are tabulated in table~\ref{table:new_datasets} and have been made publicly available at \cite{warwickem!}. In total, 76826 wavefunction have been simulated for multiple materials. To investigate ANN performance as the distribution of possible wavefunctions is restricted, we also simulated 11870 wavefunctions with smaller simulation hyperparameter upper bounds that reduce ranges by factors close to 1/4. In addition, we simulated 9644 wavefunctions for a randomly selected single material, In$_{1.7}$K$_2$Se$_8$Sn$_{2.28}$\cite{hwang2004cooling}, shown in fig.~\ref{fig:random_material}. Datasets were simulated for Kirkland potential summations in eqn.~\ref{eqn:kirkland_sum} to $n=3$, or truncated to $n=1$ terms. Truncating summations allows alternative simulation physics to be investigated.

\section{Artificial Neural Networks}\label{sec:direct_prediction}

\begin{figure}[tbh!]
\centering
\includegraphics[width=0.86\columnwidth]{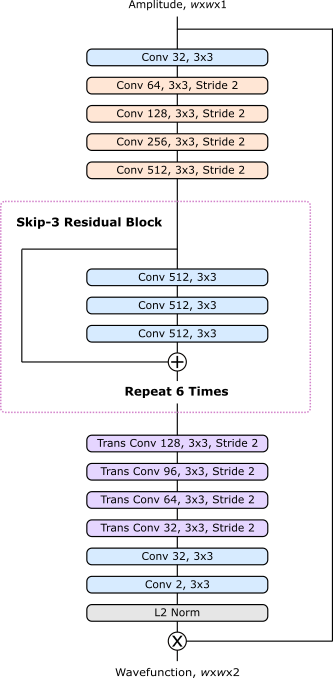}
\caption{ A convolutional neural network generates $w$$\times$$w$$\times$2 channelwise concatenations of wavefunction components from their amplitudes. Training MSEs are calculated for phase components, before multiplication by input amplitudes. }
\label{fig:generator}
\end{figure}

To reconstruct an exit wavefunction, $\psi_\text{exit}$, from its amplitude, $A$, an ANN must recover missing phase information, $\theta$. However, $\theta \in [-\infty, \infty]$, and restricting phase support to one period of the phase is complicated by cyclic periodicity. Instead, it is convenient to predict a periodic function of the phase with finite support. We use two output channels in fig.~\ref{fig:generator} to predict phase components, $\cos\theta$ and $\sin\theta$, where $\psi = A(\cos\theta +  i \sin\theta)$.

Each convolutional layer\cite{mccann2017convolutional, krizhevsky2012imagenet} is followed by batch normalization\cite{ioffe2015batch}, then activation, except the last layer where no activation is applied. Convolutional layers in residual blocks\cite{he2015deep} are ReLU\cite{nair2010rectified} activated, whereas slope 0.1 leaky ReLU\cite{maas2013rectifier} activation is used after other convolutional layers to avoid dying ReLU\cite{lu2019dying, douglas2018relu, xu2015empirical}. In denomination, channelwise L2 normalization imposes the identity $|\exp( i \theta )| \equiv 1$ after the final convolutional layer. 

\begin{figure}[tbh!]
\centering
\includegraphics[width=0.46\columnwidth]{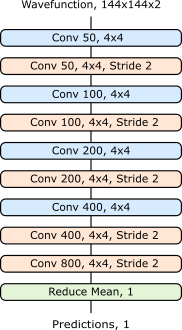}
\caption{ A discriminator predicts if wavefunction components were generated by a neural network. }
\label{fig:discriminator}
\end{figure}

In initial experiments, batch normalization was frozen halfway through training, similar to \cite{chen2018encoder}. However, scale invariance before L2 normalization resulted in numerical instability. As a result, we updated batch normalization parameters throughout training. Adding a secondary objective to impose a single output scale; such as a distance between mean L2 norms and unity, slowed training. Nevertheless, L2 normalization can be removed for generators that converge to low errors if $|\exp( i \theta )| \equiv 1$ is implicitly imposed by their loss functions.

For direct prediction, generators were trained by ADAM optimized\cite{kingma2014adam} stochastic gradient descent\cite{ruder2016overview, zou2018stochastic} for $i_\text{max} = 5 \times 10^5$ iterations to minimize adaptive learning rate clipped\cite{ede2019adaptive} (ALRC) mean squared errors (MSEs) of phase components. Training losses were calculated by multiplying MSEs by 10 and ALRC layers were initialized with first raw moment $\mu_1 = 25$, second raw moment $\mu_2 = 30$, exponential decay rates $\beta_1 = \beta_2 = 0.999$, and $n = 3$ standard deviations. We used an initial learning rate $\eta_0 = 0.002$, which was stepwise exponentially decayed\cite{ge2019step} by a factor of 0.5 every $i_\text{max}/7$ iterations, and a first moment of the momentum decay rate, $\beta_1=0.9$. 

In practice, wavefunctions with similar amplitudes may make output phase components ambiguous. As a result, a MSE trained generator may predict a weighted mean of multiple probable phase outputs, even if it understands that one pair of phase components is more likely. To overcome this limitation, we propose training a generative adversarial network\cite{goodfellow2014generative} (GAN) to predict most probable outputs. Specifically, we propose training a discriminator, $D$, in fig.~\ref{fig:discriminator} for a function, $f$, of amplitudes, and real and generated output phase components. This will enable an adversarial generator, $G$, to learn to output realistic phases in the context of their amplitudes.

There are many popular GAN loss functions and regularization mechanisms\cite{wang2019generative, dong2019towards}. Following \cite{miyato2018spectral}, we use mean squared generator, $L_G$, and discriminator, $L_D$, losses, and apply spectral normalization to the weights of every convolutional layer in the discriminator
\begin{align}
L_D &= (D(f(\psi))-1)^2 + D(f(G(|\psi|)))^2 \\
L_G &= (D(f(G(|\psi|))-1)^2,
\end{align}
where $f$ is a function that parameterizes $\psi$ as the channelwise concatenation of $\{A\cos\theta, A\sin\theta\}$. Multiplying generated phase components by inputted $A$ conditions wavefunction discrimination on $A$, ensuring that the generator learns to output physically probable $\theta$. Other parameterizations; such as the channelwise concatenation of $\{A, \cos\theta, \sin\theta\}$ could also be used. There are no biases in the discriminator.

Concatenation of conditional information to discriminator inputs and feature channels is investigated in \cite{mirza2014conditional, denton2015deep, reed2016generative, zhang2017stackgan, perarnau2016invertible, saito2017temporal, dumoulin2016adversarially, sricharan2017semi}. Projection discriminators, which calculate inner products of generator outputs and conditional embeddings, are an alternative that achieve higher performance in \cite{miyato2018cgans}. However, blind compression to an embedded representation would reduce wavefunction information, potentially limiting the quality of generated wavefunctions, and may encourage catastrophic forgetting\cite{liang2018generative}.

\afterpage{%
\begin{figure*}[tbh!]
{\centering
\includegraphics[width=\textwidth]{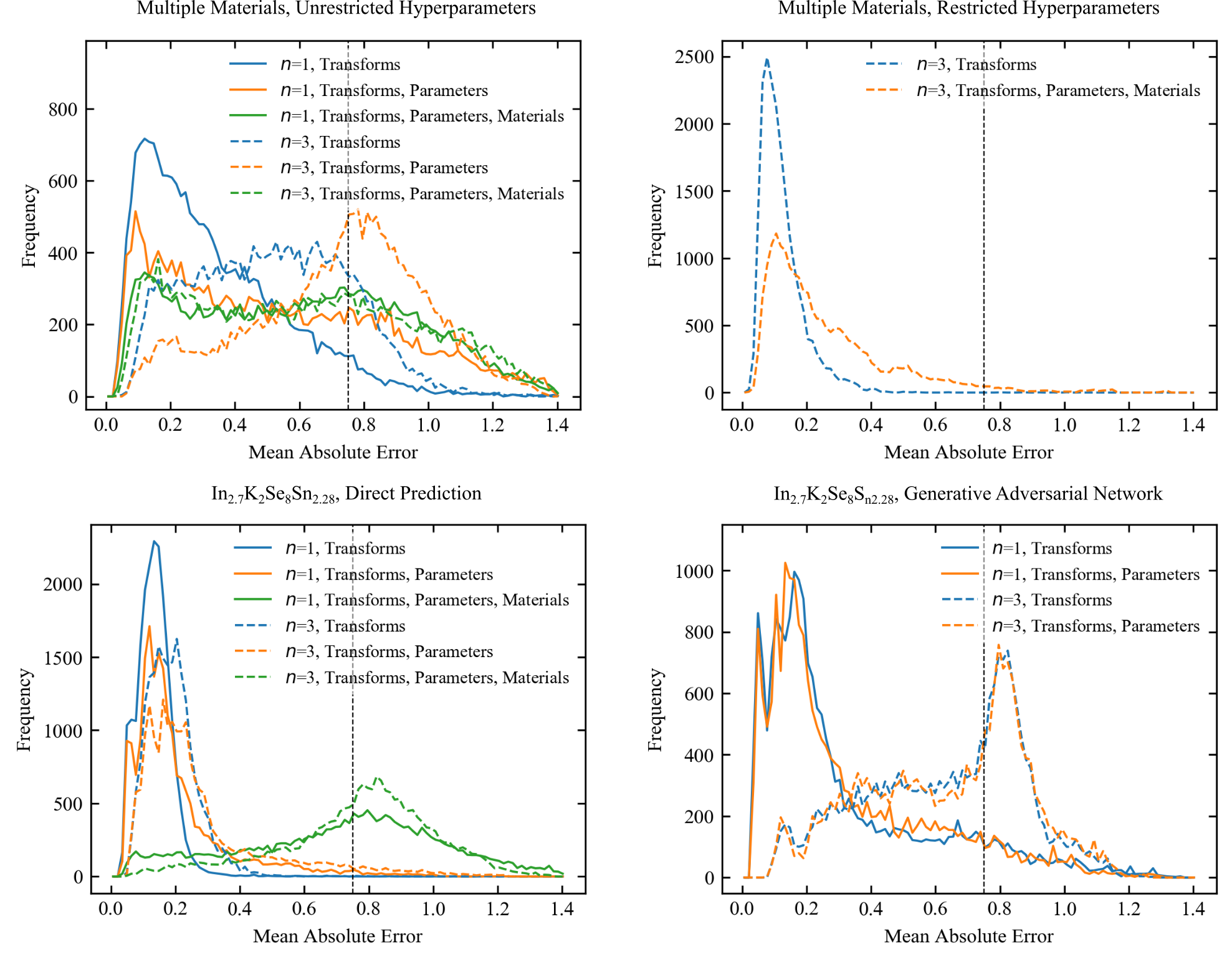}
\caption{ Frequency distributions show 19992 validation set mean absolute errors for neural networks trained to reconstruct wavefunctions simulated for multiple materials, multiple materials with restricted simulation hyperparameters, and In$_{1.7}$K$_2$Se$_8$Sn$_{2.28}$. Networks for In$_{1.7}$K$_2$Se$_8$Sn$_{2.28}$ were trained to predict phase components directly; minimising squared errors, and as part of generative adversarial networks. To demonstrate robustness to simulation physics, some validation set errors are shown for $n=1$ and $n=3$ simulation physics. We used up to three validation sets, which cumulatively quantify the ability of a network to generalize to unseen transforms; combinations of flips, rotations and translations, simulation hyperparameters; such as thickness and voltage, and materials. A vertical dashed line indicates an expected error of 0.75 for random phases, and frequencies are distributed across 100 bins. }
\label{fig:phase_results}}
\footnotesize
\vspace{\baselineskip}
\begin{tabular*}{\textwidth}{@{\extracolsep{\fill}}lccccccc}
\hline
\multicolumn{1}{c}{} & \multicolumn{1}{c}{}  & \multicolumn{2}{c}{Trans.} & \multicolumn{2}{c}{Trans., Param.} & \multicolumn{2}{c}{Trans., Param., Mater.} \\
\multicolumn{1}{c}{Training Scope} & $n$ & Mean       & Std Dev      & Mean       & \multicolumn{1}{c}{Std Dev}       & Mean      & Std Dev      \\ \hline
Multiple Materials, Unrestricted Parameters & 1 & 0.333 & 0.220 & 0.525 & 0.341 & 0.600 & 0.334 \\
In$_{1.7}$K$_2$Se$_8$Sn$_{2.28}$, MSE & 1 & 0.135 & 0.056 & 0.205 & 0.157 & 0.708 & 0.310 \\
In$_{1.7}$K$_2$Se$_8$Sn$_{2.28}$, GAN & 1 & 0.318 & 0.279 & 0.321 & 0.256 & - & - \\
\hline
Multiple Materials, Unrestricted Parameters & 3 & 0.513 & 0.234 & 0.717 & 0.271 & 0.614 & 0.344 \\
Multiple Materials, Restricted Parameters & 3 & 0.123 & 0.069 & - & - & 0.260 & 0.192 \\
In$_{1.7}$K$_2$Se$_8$Sn$_{2.28}$, MSE & 3 & 0.190 & 0.079 & 0.281 & 0.208 & 0.768 & 0.235 \\
In$_{1.7}$K$_2$Se$_8$Sn$_{2.28}$, GAN & 3 & 0.633 & 0.244 & 0.638 & 0.249 & - & - \\
\hline
Uniform Random Phases (Max Entropy) & 1, 3 & 0.750 & 0.520 & 0.750 & 0.520 & 0.750 & 0.520 \\
\hline
\end{tabular*}
\captionof{table}{ Means and standard deviations of 19992 validation set errors for unseen transforms (trans.), simulations hyperparameters (param.) and materials (mater.). All networks outperform a baseline uniform random phase generator for both $n=1$ and $n=3$ simulation physics. Dashes (-) indicate that validation set wavefunctions have not been simulated. }
\label{table:phase_results}
\end{figure*}
}

Both generator and discriminator training was ADAM optimized for $5 \times 10^5$ iterations with base learning rate $\eta_{G} = \eta_{D} = 0.0002$, and first moment of the momentum decay, $\beta_1=0.5$. To balance generator and discriminator learning, we map the discriminator learning rate to
\begin{equation}
\eta_D' = \dfrac{\eta_{D}}{1+\exp(-m(\mu_D-c))},
\end{equation}
where $\mu_D$ is the running mean discrimination for generated wavefunctions, $D(f(G(|\psi|))$, tracked by an exponential moving average with a decay rate of 0.99, and $m=20$ and $c=0.5$ linearly transform $\mu_D$. 

To augment training data, we selected random $w$$\times$$w$ crops from 320$\times$320 wavefunctions. Each crop was then subject to random combination of flips and $\pi$/2 rad rotations to augment our datasets by a factor of eight. We chose wavefunction size $w = 224$ for direct prediction and $w = 144$ for GANs, where $w$ is smaller for GANs as discriminators add to GPU memory requirements. ANNs were trained with a batch size of 24.

\section{Experiments}

In this section, we investigate phase recovery with ANNs as the distribution of wavefunctions is restricted. To directly predict $\theta$ for $A$, we trained ANNs for multiple materials, multiple materials with restricted simulation hyperparameters, and In$_{1.7}$K$_2$Se$_8$Sn$_{2.28}$. We also trained a GAN for In$_{1.7}$K$_2$Se$_8$Sn$_{2.28}$ wavefunctions. Experiments are repeated with the summation in eqn.~\ref{eqn:kirkland_sum} truncated from $n=3$ to $n=1$, to demonstrate robustness to simulation physics.

Distributions of generated phase component mean absolute errors (MAEs) for sets of 19992 validation examples are shown in fig.~\ref{fig:phase_results}, and moments are tabulated in table~\ref{table:phase_results}. We used up to three validation sets, which cumulatively quantify the ability of a network to generalize to unseen transforms; combinations of flips, rotations and translations, simulation hyperparameters; such as thickness and voltage, and materials. In comparison, the expected error of the $n$th moment of phase components, $\text{E}[|G(|\psi|) - f(\theta)|^n]$, where $g \in \{\cos, \sin\}$, for uniform random predictions, $x \sim  U(-1, 1)$, and uniformly distributed phases, $\theta \sim  U(-\pi, \pi)$, is 
\begin{align}
\text{E}[|x - g(\theta)|^n] &= \int\limits_{-1}^1\int\limits_{-\pi}^\pi \rho(x)\rho(\theta)|x - g(\theta)|^n \diff{\theta}\diff{x},
\end{align}
where $\rho(\theta) = 1/2\pi$ and $\rho(x) = 1/2$ are uniform probability density functions for $\theta$ and $x$, respectively. The first two moments are $\text{E}[|x - g(\theta)|] = 3/4$ and $\text{E}[|x - g(\theta)|^2] = 5/6$; making the expected standard deviation 0.520.

All ANN MAEs have lower means and standard deviations than a baseline random phase generator, except a In$_{1.7}$K$_2$Se$_8$Sn$_{2.28}$ generator applied to other materials. ANNs do not have prior understanding of propagation equations or dynamics. As a result, experiments demonstrate that ANNs are able to develop and leverage a physical understanding to recover $\theta$. ANNs are trained for Kirkland potential summations in eqn.~\ref{eqn:kirkland_sum} to $n=3$ and $n=1$ terms, demonstrating a robustness to simulation physics. Success with different simulation physics motivates the development of ANNs for real physics; approximated by $n=3$ simulation physics.

Validation set MAEs increase as wavefunction restrictions are cumulatively reduced from unseen transforms used for data augmentation during training, to unseen simulation parameters, and unseen materials. For example, MAEs are 0.600 and 0.614 for ANNs trained for multiple materials, increasing to 0.708 and 0.768 for ANNs trained for In$_{1.7}$K$_2$Se$_8$Sn$_{2.28}$. This shows that MAEs increase for materials an ANN is unfamiliar with, approaching MAEs of 0.75 expected for a uniform random phase generator where there is no familiarity.

Wavefunctions are insufficiently restricted for multiple materials. Validation MAEs of 0.333 and 0.513 for unseen transforms diverge to 0.600 and 0.614 for unseen simuation hyperparamaters and materials. In addition, a peak near 0.15 decreases, and MAE density around 0.75 increases. Taken together, this indicates that multiple material ANNs are able to recognise and generalize to some wavefunctions; however, their ability to generalize is limited. Further, frequency distribution tails exceed 0.75 for all validation sets. This may indicate that the generator struggles with material and simulation or hyperparameter combinations that produce wavefunctions with unusual or unpalatable characteristics. However, we believe the tail is mainly caused by combinations that produce different wavefunctions with similar amplitudes.

\begin{figure}[tbp!]
\centering
\includegraphics[width=0.97\columnwidth]{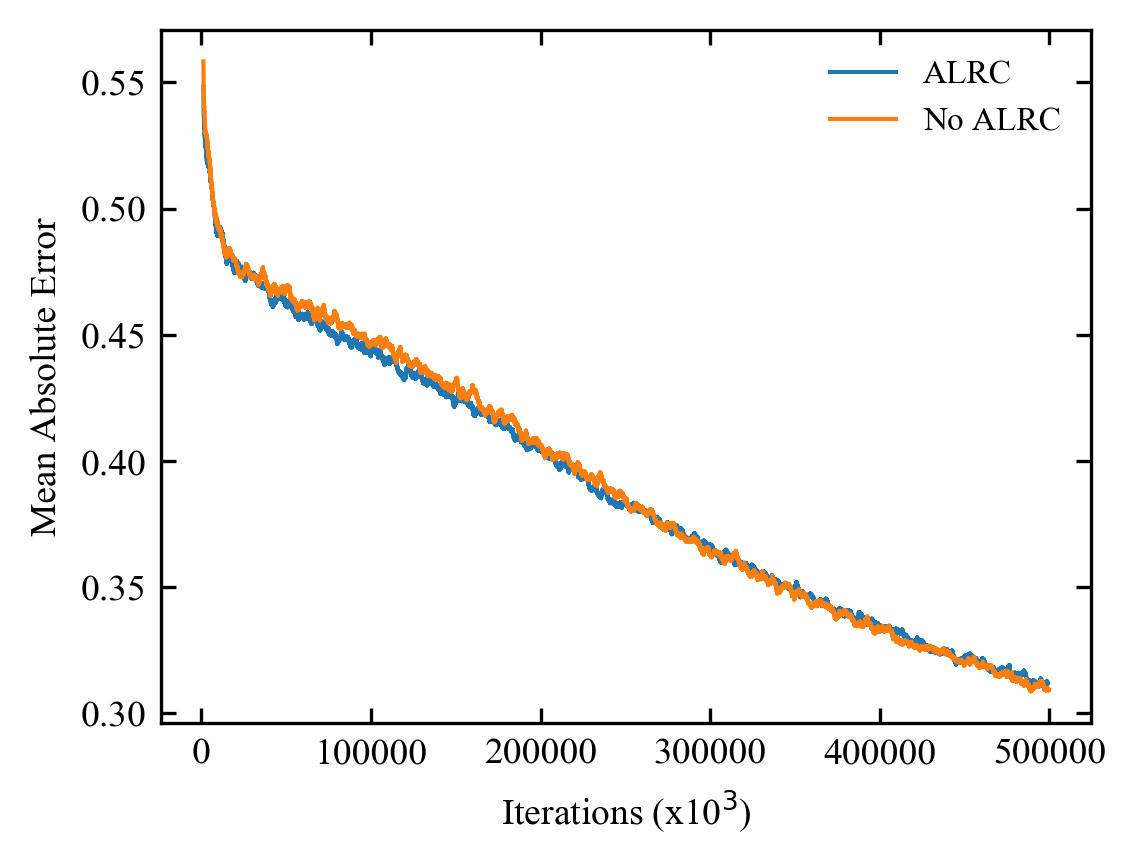}
\caption{ Training mean absolute errors are similar with and without adaptive learning rate clipping (ALRC). Learning curves are 2500 iteration boxcar averaged. }
\label{fig:alrc_vs_no_alrc}
\end{figure}

Validation divergence decreases as the distribution of wavefunctions is restricted. For example, frequency distributions have almost no tail beyond 0.75 for simulation hyperparameter ranges reduced by factors close to 1/4. Validation divergence is also reduced by training for In$_{1.7}$K$_2$Se$_8$Sn$_{2.28}$, a single material. Restricting the distribution of wavefunctions is an essential part of one-shot wavefunction reconstruction, otherwise there is an infinite number of possible $\theta$ for $A$.

To investigate an approach to reduce prediction weighting for $A$ with a range of probable $\theta$, we trained GANs for In$_{1.7}$K$_2$Se$_8$Sn$_{2.28}$. Training as part of a GAN acts as a regularization mechanism, lowering validation divergence. However, a GAN requires a powerful discriminator to understand the distribution of possible wavefunctions and can be difficult to train. In particular, $n=3$ wavefunctions have lower local spatial correlation than $n=1$ wavefunctions at our simulation resolution, which made it more difficult for our $n=3$ GAN to learn.

\begin{figure*}[tbp!]
\centering
\footnotesize
\includegraphics[width=\textwidth]{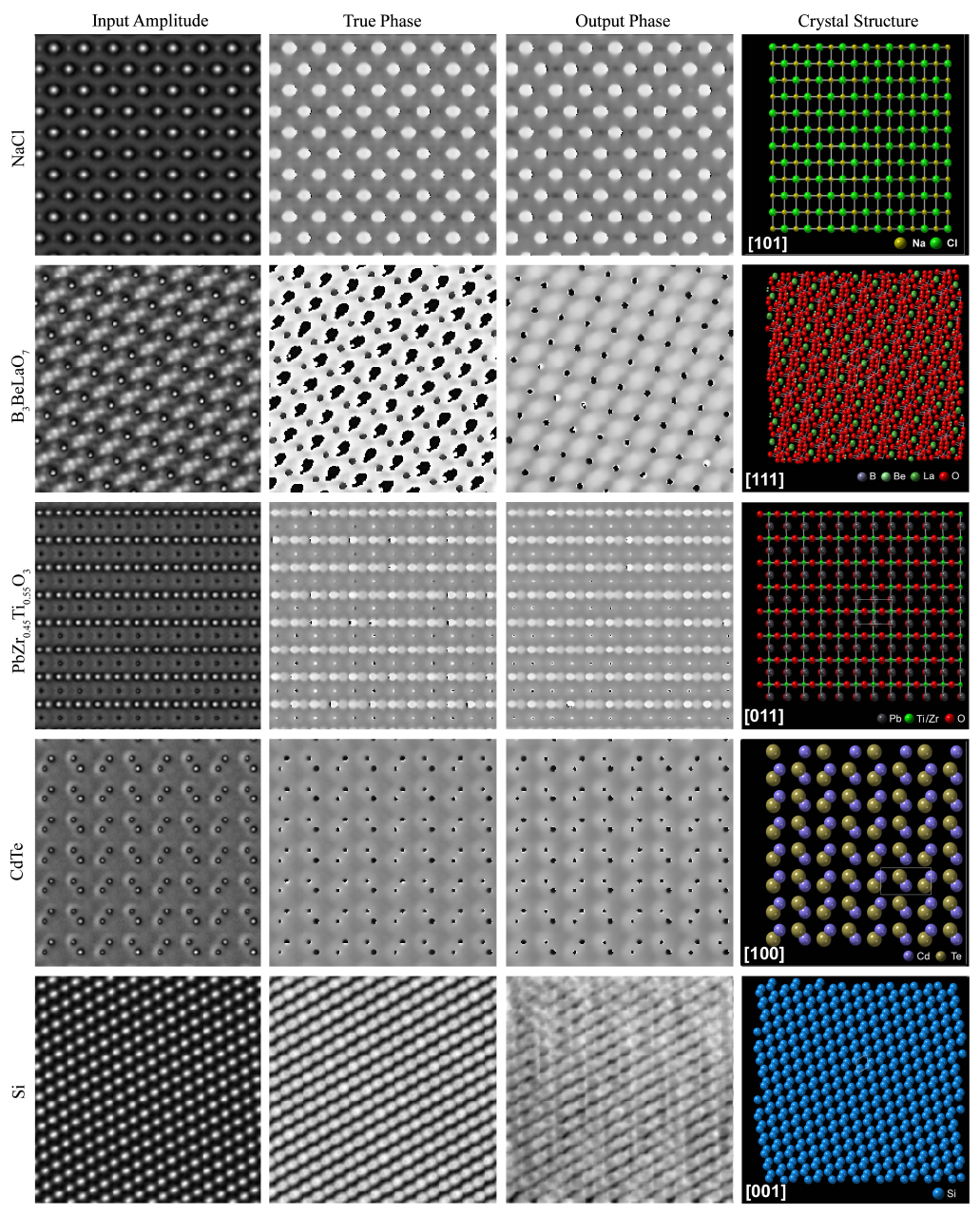}
\caption{ Exit wavefunction reconstruction for unseen  NaCl, B$_3$BeLaO$_7$, PbZr$_{0.45}$Ti$_{0.55}$0$_3$, CdTe, and Si input amplitudes, and corresponding crystal structures. Phases in $[-\pi, \pi)$ rad are depicted on a linear greycale from black to white, and show that output phases are close to true phases. Wavefunctions are cyclically periodic functions of phase so distances between black and white pixels are small. Si is a failure case where phase information is not accurately recovered. Miller indices label projection directions. }
\label{fig:unseen_materials}
\end{figure*}

Training loss distributions have tails with high losses. As a result, we used ALRC to limit high errors. A comparison of training with and without ALRC is in fig.~\ref{fig:alrc_vs_no_alrc}. Validation MAEs for unseen materials have mean 0.600 and standard deviation 0.334 with ALRC, and mean 0.602 and standard deviation 0.338 without ALRC. Differences between validation MAEs is insignificant, so ALRC is not helping for training with batch size 24. This behavior is in-line with results in the ALRC paper\cite{ede2019adaptive}, which shows that ALRC becomes less effective as batch size increases. Nevertheless, ALRC may be help lower error if generators are trained with smaller batch sizes. In particular, if the wavefunction distribution is restricted so errors are low, removing the need for L2 normalization at the end of the generator, and therefore decreasing dependence on batch normalization.

Examples of ANN phase recovery are shown in fig.~\ref{fig:unseen_materials} alongside crystal structures highlighting the structural information producing exit wavefunctions. Results are for unseen materials and an ANN trained for multiple materials with restricted simulation hyperparameters. Wavefunctions are presented for NaCl\cite{abrahams1965accuracy} and elemental Si as they are simple materials with widely recognised structures. Other materials belong to classes that are widely investigated: B$_3$BeLaO$_7$\cite{yan2013labeb} is a non-linear optical crystal, PbZr$_{0.45}$Ti$_{0.55}$0$_3$\cite{idemoto2004relation} is ferroelectric used in ultrasonic transducers\cite{chen2018pzt} and ceramic capacitors\cite{hikam2018pyroelectric}, and CdTe is a semiconductor used in solar cells\cite{burst2016cdte}. The Si example is also included as typical failure case for unfamiliar examples. In this case, possibly because the Si crystal structure is unusually simple. Additional sheets of example input phases, generated phases, and true phases for each ANN will be provided as supplementary information with the published version of this preprint.

\section{Discussion}

This paper describes an initial investigation into CTEM one-shot exit wavefunction reconstruction with deep learning, and is intended to be a starting point for future research. We expect that ANN architecture and learning policy can be substantially improved; possibly with AdaNet\cite{weill2019adanet}, Ludwig\cite{molino2019ludwig}, or other automatic machine learning\cite{he2019automl} algorithms, and we encourage further investigation. In this spirit, all of our source code\cite{one-shot_repo} (based on TensorFlow\cite{abadi2016tensorflow}), clTEM simulation software\cite{clTEM_repo}, and new wavefunction datasets\cite{warwickem!} have been made publicly available. Training for each network was stopped after a few days on an Nvidia 1080 Ti GPU, and losses were still decreasing. As a result, this paper presents lower bounds for performance.

To demonstrate robustness to simulation physics, Kirkland potential summations in eqn.~\ref{eqn:kirkland_sum} were calculated with $n=3$, or truncated to $n=1$ terms, for different datasets. For further simulations, compiled clTEM versions with $n=1$ and $n=3$ have been included in our project repository\cite{one-shot_repo}. Source code for clTEM is also available with separate pre-releases\cite{clTEM_repo}. Summations with $n=3$ approximate experimental physics, whereas $n=1$ is for an alternative universe with different atom potentials.

Our experiments do not include aberrations or detector noise. This restricts the distribution of wavefunctions and makes it easier for ANNs to learn. However, distributions of wavefunctions were less restricted than possible in practice, and ANNs can remove noise\cite{ede2019improving}. As a result, we expect one-shot exit wavefunction to be applicable to experimental images. A good starting point for future research may be materials where the distribution of wavefunctions is naturally restricted. For example, graphene\cite{wang2019characterization} and other two-dimensional materials\cite{mendes2019electron}, select crystals at atomic resolution\cite{zhang2018atomic}, or classified images; such as biological specimens\cite{lakshman2019application, ogawa2019transmission} after similar preparation.

Information about materials, expected ranges of simulation hyperparameters, and other metadata was not input to ANNs. However, this variable information is readily available and could restrict the distribution of wavefunctions; improving ANN performance. Subsequently, we suggest that metadata embedded by an ANN could be used to modulate information transfer through a convolutional neural network by conditional batch normalization\cite{perez2017learning}. However, metadata is typically high-dimensional, so this may be impractical beyond individual applications. 

By default, large amounts of metadata is saved to Digital Micrograph image files (e.g. dm3 and dm4) created by Gatan Microscopy Suite\cite{gms_webpage} software. Metadata can also be saved to TIFFs\cite{adobe1992tiff} or other image formats preferred by electron microscopists using different software. In practice, most of this metadata describes microscope settings; such as voltage and magnification, and may not be sufficient to restrict the distribution of wavefunctions. Nevertheless, most file formats support the addition of extra metadata that is readily known to experimenters. Example information may include estimates for stoichiometry, specimen thickness, zone axis, temperature, the microscope and its likely aberration range, and phenomena exhibited by materials in scientific literature. ANNs have been developed to embed scientific literature\cite{tshitoyan2019unsupervised}, so we expect that it will become possible to include additional metadata as a lay description.

In this paper, ANNs are trained to reconstruct $\psi$ from $A$, and therefore follow a history of successful deep learning applications to accelerated quantum mechanics\cite{beach2018qucumber, carleo2019netket}. In contrast, experimental holograms are integrated over detector supports. Although probability density, $|\psi(\bar{S})|^2$, at the mean support, $\bar{S}$, can be factored outside the integral of eqn.~\ref{eqn:intensity_integral} if spatial variation is small, $\nabla \chi \rightarrow 0$, and $S$ is effectively invariant,
\begin{equation}\label{eqn:measure_approx}
I(S) \approx |\psi(\bar{S})|^2 \int\limits_{s\in S} \diff{s},
\end{equation}
these restrictions are unrealistic. In practice, we do not think the distinction is important as ANNs have learned to recover optical $\theta$ from $I$\cite{rivenson2018phase}.

To discourage ANNs from gaming their loss functions by predicting an average of probable phase components, we propose training GANs. However, GANs are difficult to train\cite{salimans2016improved, liang2018generative}, and GAN training can take longer than with MSEs. For example, our validation set GAN MAEs are lower than for MSE training after $5 \times 10^5$ iterations. We also found that GAN performance can be much lower for some wavefunctions; such as those with low local spatial correlation. High performance for large wavefunctions also requires powerful discriminators; such as \cite{brock2018large}, to understand their distribution. 

Overall, we expect GANs to become less useful the more a distribution of wavefunctions is restricted. As the distribution becomes more restricted, a smaller portion of the distribution has similar amplitudes with substantially different phases. In part, we expect this effect already lowers MAEs as distributions are restricted. Another contribution is restricted physics; which makes networks less reliant on identifying features. As a result, we expect the main use of GANs in phase recovery to be improving wavefunction realism. 

\section{Conclusions}

We have simulated five new datasets containing 98340 CTEM exit wavefunctions with clTEM. The datasets have been used to train ANNs to reconstruct wavefunctions from single images. In this initial investigation, we found that ANN performance improves as the distribution of wavefunctions is restricted. One-shot exit wavefunction reconstruction overcomes the limitations of aberration series reconstruction and holography: it is live, does not require experimental equipment, and can be applied as a post-processing step indefinitely after an image is taken. We expect our results to be generalizable to other types of electron microscopy.

\section{Supplementary Information}

This work is intended to establish starting points to be improved on by future research. In this spirit, our new datasets\cite{warwickem!}, clTEM simulation software\cite{clTEM_repo}, and source code with links to pre-trained models\cite{one-shot_repo} has been made publicly available. 

In appendices, we build on Abbe's theory of wave optics to propose a new approach to phase recovery with deep learning. The idea is that wavefunctions could be learned from large datasets of single images; avoiding the difficulty and expense of collecting experimental wavefunctions. Nevertheless, we also introduce a new dataset containing 1000 512$\times$512 experimental focal series. In addition, a supplementary document will be provided with the published version of this preprint with sheets of example input amplitudes, output phases, and true phases for every ANN featured in this paper. 

\bibliography{bibliography}

\begin{thebibliography}{10}
\expandafter\ifx\csname url\endcsname\relax
  \def\url#1{\texttt{#1}}\fi
\expandafter\ifx\csname urlprefix\endcsname\relax\def\urlprefix{URL }\fi
\expandafter\ifx\csname href\endcsname\relax
  \def\href#1#2{#2} \def\path#1{#1}\fi

\bibitem{lehmann2002tutorial}
M.~Lehmann, H.~Lichte, Tutorial on off-axis electron holography, Microscopy and
  Microanalysis 8~(6) (2002) 447--466.

\bibitem{frabboni2007young}
S.~Frabboni, G.~C. Gazzadi, G.~Pozzi, Young’s double-slit interference
  experiment with electrons, American Journal of Physics 75~(11) (2007)
  1053--1055.

\bibitem{matteucci1998experiment}
G.~Matteucci, C.~Beeli, An experiment on electron wave--particle duality
  including a {P}lanck constant measurement, American Journal of Physics
  66~(12) (1998) 1055--1059.

\bibitem{lentzen2000reconstruction}
M.~Lentzen, K.~Urban, Reconstruction of the projected crystal potential in
  transmission electron microscopy by means of a maximum-likelihood refinement
  algorithm, Acta Crystallographica Section A: Foundations of Crystallography
  56~(3) (2000) 235--247.

\bibitem{auslender2019measuring}
A.~Auslender, M.~Halabi, G.~Levi, O.~Di{\'e}guez, A.~Kohn, Measuring the mean
  inner potential of {Al}$_2${O}$_3$ sapphire using off-axis electron
  holography, Ultramicroscopy 198 (2019) 18--25.

\bibitem{tonomura1987applications}
A.~Tonomura, Applications of electron holography, Reviews of modern physics
  59~(3) (1987) 639.

\bibitem{fu1991correction}
Q.~Fu, H.~Lichte, E.~V{\"o}lkl, Correction of aberrations of an electron
  microscope by means of electron holography, Physical review letters 67~(17)
  (1991) 2319.

\bibitem{mccartney1994absolute}
M.~McCartney, M.~Gajdardziska-Josifovska, Absolute measurement of normalized
  thickness, $t/\lambda_i$, from off-axis electron holography, Ultramicroscopy
  53~(3) (1994) 283--289.

\bibitem{park2014observation}
H.~S. Park, X.~Yu, S.~Aizawa, T.~Tanigaki, T.~Akashi, Y.~Takahashi, T.~Matsuda,
  N.~Kanazawa, Y.~Onose, D.~Shindo, et~al., Observation of the magnetic flux
  and three-dimensional structure of skyrmion lattices by electron holography,
  Nature nanotechnology 9~(5) (2014) 337.

\bibitem{trippoff}
R.~E. Dunin-Borkowski, T.~Kasama, A.~Wei, S.~L. Tripp, M.~J. H{\"y}tch,
  E.~Snoeck, R.~J. Harrison, A.~Putnis, Off-axis electron holography of
  magnetic nanowires and chains, rings, and planar arrays of magnetic
  nanoparticles, Microscopy research and technique 64~(5-6) (2004) 390--402.

\bibitem{mcmullan2016direct}
G.~McMullan, A.~Faruqi, R.~Henderson, Direct electron detectors, in: Methods in
  enzymology, Vol. 579, Elsevier, 2016, pp. 1--17.

\bibitem{mcmullan2009detective}
G.~McMullan, S.~Chen, R.~Henderson, A.~Faruqi, Detective quantum efficiency of
  electron area detectors in electron microscopy, Ultramicroscopy 109~(9)
  (2009) 1126--1143.

\bibitem{carter2016transmission}
C.~B. Carter, D.~B. Williams, Transmission electron microscopy: Diffraction,
  imaging, and spectrometry, Springer, 2016.

\bibitem{pennycook2011scanning}
S.~J. Pennycook, P.~D. Nellist, Scanning transmission electron microscopy:
  imaging and analysis, Springer Science \& Business Media, 2011.

\bibitem{goldstein2017scanning}
J.~I. Goldstein, D.~E. Newbury, J.~R. Michael, N.~W. Ritchie, J.~H.~J. Scott,
  D.~C. Joy, Scanning electron microscopy and X-ray microanalysis, Springer,
  2017.

\bibitem{kohler1981abbe}
H.~K{\"o}hler, On {A}bbe's theory of image formation in the microscope, Optica
  Acta: International Journal of Optics 28~(12) (1981) 1691--1701.

\bibitem{lubk2016fundamentals}
A.~Lubk, K.~Vogel, D.~Wolf, J.~Krehl, F.~R{\"o}der, L.~Clark, G.~Guzzinati,
  J.~Verbeeck, Fundamentals of focal series inline electron holography, in:
  Advances in imaging and electron physics, Vol. 197, Elsevier, 2016, pp.
  105--147.

\bibitem{koch2010off}
C.~T. Koch, A.~Lubk, Off-axis and inline electron holography: A quantitative
  comparison, Ultramicroscopy 110~(5) (2010) 460--471.

\bibitem{koch2014towards}
C.~T. Koch, Towards full-resolution inline electron holography, Micron 63
  (2014) 69--75.

\bibitem{haigh2013recording}
S.~Haigh, B.~Jiang, D.~Alloyeau, C.~Kisielowski, A.~Kirkland, Recording low and
  high spatial frequencies in exit wave reconstructions, Ultramicroscopy 133
  (2013) 26--34.

\bibitem{ozsoy2014hybridization}
C.~Ozsoy-Keskinbora, C.~Boothroyd, R.~Dunin-Borkowski, P.~Van~Aken, C.~Koch,
  Hybridization approach to in-line and off-axis (electron) holography for
  superior resolution and phase sensitivity, Scientific Reports 4 (2014) 7020.

\bibitem{ruskin2013quantitative}
R.~S. Ruskin, Z.~Yu, N.~Grigorieff, Quantitative characterization of electron
  detectors for transmission electron microscopy, Journal of structural biology
  184~(3) (2013) 385--393.

\bibitem{rivenson2018phase}
Y.~Rivenson, Y.~Zhang, H.~G{\"u}nayd{\i}n, D.~Teng, A.~Ozcan, Phase recovery
  and holographic image reconstruction using deep learning in neural networks,
  Light: Science \& Applications 7~(2) (2018) 17141.

\bibitem{morgan2011direct}
A.~Morgan, A.~Martin, A.~D'Alfonso, C.~Putkunz, L.~Allen, Direct exit-wave
  reconstruction from a single defocused image, Ultramicroscopy 111~(9-10)
  (2011) 1455--1460.

\bibitem{martin2008direct}
A.~Martin, L.~Allen, Direct retrieval of a complex wave from its diffraction
  pattern, Optics Communications 281~(20) (2008) 5114--5121.

\bibitem{warwickem!}
J.~M. Ede, J.~P.~P. Peters, R.~Beanland, Warwick electron microscopy datasets,
  online:
  \url{https://warwick.ac.uk/fac/sci/physics/research/condensedmatt/microscopy/research/machinelearning}
  (2019).

\bibitem{clTEM_repo}
J.~P.~P. Peters, M.~A. Dyson, cl{TEM}, online:
  \url{https://github.com/JJPPeters/clTEM} (2019).

\bibitem{dyson2014advances}
M.~A. Dyson, Advances in computational methods for transmission electron
  microscopy simulation and image processing, Ph.D. thesis, University of
  Warwick (2014).

\bibitem{hall1991crystallographic}
S.~R. Hall, F.~H. Allen, I.~D. Brown, The crystallographic information file
  ({CIF}): a new standard archive file for crystallography, Acta
  Crystallographica Section A: Foundations of Crystallography 47~(6) (1991)
  655--685.

\bibitem{Quiros2018}
M.~Quir{\'{o}}s, S.~Gra{\v{z}}ulis, S.~Girdzijauskait{\.{e}}, A.~Merkys,
  A.~Vaitkus, Using {SMILES} strings for the description of chemical
  connectivity in the {C}rystallography {O}pen {D}atabase, Journal of
  Cheminformatics 10~(1) (May 2018).
\newblock \href {https://doi.org/10.1186/s13321-018-0279-6}
  {\path{doi:10.1186/s13321-018-0279-6}}.

\bibitem{Merkys2016}
A.~Merkys, A.~Vaitkus, J.~Butkus, M.~Okuli{\v{c}}-Kazarinas, V.~Kairys,
  S.~Gra{\v{z}}ulis,
  \href{http://dx.doi.org/10.1107/S1600576715022396}{{COD}::{CIF}::{P}arser: an
  error-correcting {CIF} parser for the {P}erl language}, Journal of Applied
  Crystallography 49~(1) (Feb 2016).
\newblock \href {https://doi.org/10.1107/S1600576715022396}
  {\path{doi:10.1107/S1600576715022396}}.
\newline\urlprefix\url{http://dx.doi.org/10.1107/S1600576715022396}

\bibitem{Grazulis2015}
S.~Gra{\v{z}}ulis, A.~Merkys, A.~Vaitkus, M.~Okuli{\v{c}}-Kazarinas,
  \href{http://dx.doi.org/10.1107/S1600576714025904}{Computing stoichiometric
  molecular composition from crystal structures}, Journal of Applied
  Crystallography 48~(1) (2015) 85--91.
\newblock \href {https://doi.org/10.1107/S1600576714025904}
  {\path{doi:10.1107/S1600576714025904}}.
\newline\urlprefix\url{http://dx.doi.org/10.1107/S1600576714025904}

\bibitem{Grazulis2012}
S.~Gra{\v{z}}ulis, A.~Da{\v{s}}kevi{\v{c}}, A.~Merkys, D.~Chateigner,
  L.~Lutterotti, M.~Quir{\'o}s, N.~R. Serebryanaya, P.~Moeck, R.~T. Downs,
  A.~Le~Bail,
  \href{http://nar.oxfordjournals.org/content/40/D1/D420.abstract}{{C}rystallography
  {O}pen {D}atabase ({COD}): an open-access collection of crystal structures
  and platform for world-wide collaboration}, Nucleic Acids Research 40~(D1)
  (2012) D420--D427.
\newblock \href
  {http://arxiv.org/abs/http://nar.oxfordjournals.org/content/40/D1/D420.full.pdf+html}
  {\path{arXiv:http://nar.oxfordjournals.org/content/40/D1/D420.full.pdf+html}},
  \href {https://doi.org/10.1093/nar/gkr900} {\path{doi:10.1093/nar/gkr900}}.
\newline\urlprefix\url{http://nar.oxfordjournals.org/content/40/D1/D420.abstract}

\bibitem{Grazulis2009}
S.~Gra{\v{z}}ulis, D.~Chateigner, R.~T. Downs, A.~F.~T. Yokochi, M.~Quir{\'o}s,
  L.~Lutterotti, E.~Manakova, J.~Butkus, P.~Moeck, A.~Le~Bail,
  \href{http://dx.doi.org/10.1107/S0021889809016690}{{{C}rystallography {O}pen
  {D}atabase {--} an open-access collection of crystal structures}}, Journal of
  Applied Crystallography 42~(4) (2009) 726--729.
\newblock \href {https://doi.org/10.1107/S0021889809016690}
  {\path{doi:10.1107/S0021889809016690}}.
\newline\urlprefix\url{http://dx.doi.org/10.1107/S0021889809016690}

\bibitem{Downs2003}
R.~T. Downs, M.~Hall-Wallace, The {A}merican {M}ineralogist crystal structure
  database, American Mineralogist 88 (2003) 247--250.

\bibitem{kirkland2010advanced}
E.~J. Kirkland, Advanced computing in electron microscopy, Springer Science \&
  Business Media, 2010.

\bibitem{stone2010opencl}
J.~E. Stone, D.~Gohara, G.~Shi, {OpenCL}: A parallel programming standard for
  heterogeneous computing systems, Computing in science \& engineering 12~(3)
  (2010) 66.

\bibitem{moreland2003fft}
K.~Moreland, E.~Angel, The {FFT} on a {GPU}, in: Proceedings of the ACM
  SIGGRAPH/EUROGRAPHICS conference on Graphics hardware, Eurographics
  Association, 2003, pp. 112--119.

\bibitem{abramowitz1964handbook}
M.~Abramowitz, I.~A. Stegun, Handbook of mathematical functions. 1965 (1964).

\bibitem{hwang2004cooling}
S.-J. Hwang, R.~G. Iyer, P.~N. Trikalitis, A.~G. Ogden, M.~G. Kanatzidis,
  Cooling of melts: Kinetic stabilization and polymorphic transitions in the
  {K}{In}{Sn}{Se}$_4$ system, Inorganic chemistry 43~(7) (2004) 2237--2239.

\bibitem{mccann2017convolutional}
M.~T. McCann, K.~H. Jin, M.~Unser, Convolutional neural networks for inverse
  problems in imaging: A review, IEEE Signal Processing Magazine 34~(6) (2017)
  85--95.

\bibitem{krizhevsky2012imagenet}
A.~Krizhevsky, I.~Sutskever, G.~E. Hinton, {ImageNet} classification with deep
  convolutional neural networks, in: Advances in neural information processing
  systems, 2012, pp. 1097--1105.

\bibitem{ioffe2015batch}
S.~Ioffe, C.~Szegedy, Batch normalization: Accelerating deep network training
  by reducing internal covariate shift, arXiv preprint arXiv:1502.03167 (2015).

\bibitem{he2015deep}
K.~He, X.~Zhang, S.~Ren, J.~Sun, Deep residual learning for image recognition.
  {CoRR} abs/1512.03385 (2015).

\bibitem{nair2010rectified}
V.~Nair, G.~E. Hinton, Rectified linear units improve restricted {B}oltzmann
  machines, in: Proceedings of the 27th international conference on machine
  learning (ICML-10), 2010, pp. 807--814.

\bibitem{maas2013rectifier}
A.~L. Maas, A.~Y. Hannun, A.~Y. Ng, Rectifier nonlinearities improve neural
  network acoustic models, in: Proc. ICML, Vol.~30, 2013, p.~3.

\bibitem{lu2019dying}
L.~Lu, Y.~Shin, Y.~Su, G.~E. Karniadakis, Dying {ReLU} and initialization:
  Theory and numerical examples, arXiv preprint arXiv:1903.06733 (2019).

\bibitem{douglas2018relu}
S.~C. Douglas, J.~Yu, Why {ReLU} units sometimes die: Analysis of single-unit
  error backpropagation in neural networks, in: 2018 52nd Asilomar Conference
  on Signals, Systems, and Computers, IEEE, 2018, pp. 864--868.

\bibitem{xu2015empirical}
B.~Xu, N.~Wang, T.~Chen, M.~Li, Empirical evaluation of rectified activations
  in convolutional network, arXiv preprint arXiv:1505.00853 (2015).

\bibitem{chen2018encoder}
L.-C. Chen, Y.~Zhu, G.~Papandreou, F.~Schroff, H.~Adam, Encoder-decoder with
  atrous separable convolution for semantic image segmentation, in: Proceedings
  of the European conference on computer vision (ECCV), 2018, pp. 801--818.

\bibitem{kingma2014adam}
D.~P. Kingma, J.~Ba, {ADAM}: A method for stochastic optimization, arXiv
  preprint arXiv:1412.6980 (2014).

\bibitem{ruder2016overview}
S.~Ruder, An overview of gradient descent optimization algorithms, arXiv
  preprint arXiv:1609.04747 (2016).

\bibitem{zou2018stochastic}
D.~Zou, Y.~Cao, D.~Zhou, Q.~Gu, Stochastic gradient descent optimizes
  over-parameterized deep relu networks, arXiv preprint arXiv:1811.08888
  (2018).

\bibitem{ede2019adaptive}
J.~M. Ede, R.~Beanland, Adaptive learning rate clipping stabilizes learning,
  arXiv preprint arXiv:1906.09060 (2019).

\bibitem{ge2019step}
R.~Ge, S.~M. Kakade, R.~Kidambi, P.~Netrapalli, The step decay schedule: A near
  optimal, geometrically decaying learning rate procedure, arXiv preprint
  arXiv:1904.12838 (2019).

\bibitem{goodfellow2014generative}
I.~Goodfellow, J.~Pouget-Abadie, M.~Mirza, B.~Xu, D.~Warde-Farley, S.~Ozair,
  A.~Courville, Y.~Bengio, Generative adversarial nets, in: Advances in neural
  information processing systems, 2014, pp. 2672--2680.

\bibitem{wang2019generative}
Z.~Wang, Q.~She, T.~E. Ward, Generative adversarial networks: A survey and
  taxonomy, arXiv preprint arXiv:1906.01529 (2019).

\bibitem{dong2019towards}
H.-W. Dong, Y.-H. Yang, Towards a deeper understanding of adversarial losses,
  arXiv preprint arXiv:1901.08753 (2019).

\bibitem{miyato2018spectral}
T.~Miyato, T.~Kataoka, M.~Koyama, Y.~Yoshida, Spectral normalization for
  generative adversarial networks, arXiv preprint arXiv:1802.05957 (2018).

\bibitem{mirza2014conditional}
M.~Mirza, S.~Osindero, Conditional generative adversarial nets, arXiv preprint
  arXiv:1411.1784 (2014).

\bibitem{denton2015deep}
E.~L. Denton, S.~Chintala, R.~Fergus, et~al., Deep generative image models
  using a {L}aplacian pyramid of adversarial networks, in: Advances in neural
  information processing systems, 2015, pp. 1486--1494.

\bibitem{reed2016generative}
S.~Reed, Z.~Akata, X.~Yan, L.~Logeswaran, B.~Schiele, H.~Lee, Generative
  adversarial text to image synthesis, arXiv preprint arXiv:1605.05396 (2016).

\bibitem{zhang2017stackgan}
H.~Zhang, T.~Xu, H.~Li, S.~Zhang, X.~Wang, X.~Huang, D.~N. Metaxas, Stackgan:
  Text to photo-realistic image synthesis with stacked generative adversarial
  networks, in: Proceedings of the IEEE International Conference on Computer
  Vision, 2017, pp. 5907--5915.

\bibitem{perarnau2016invertible}
G.~Perarnau, J.~Van De~Weijer, B.~Raducanu, J.~M. {\'A}lvarez, Invertible
  conditional {GAN}s for image editing, arXiv preprint arXiv:1611.06355 (2016).

\bibitem{saito2017temporal}
M.~Saito, E.~Matsumoto, S.~Saito, Temporal generative adversarial nets with
  singular value clipping, in: Proceedings of the IEEE International Conference
  on Computer Vision, 2017, pp. 2830--2839.

\bibitem{dumoulin2016adversarially}
V.~Dumoulin, I.~Belghazi, B.~Poole, O.~Mastropietro, A.~Lamb, M.~Arjovsky,
  A.~Courville, Adversarially learned inference, arXiv preprint
  arXiv:1606.00704 (2016).

\bibitem{sricharan2017semi}
K.~Sricharan, R.~Bala, M.~Shreve, H.~Ding, K.~Saketh, J.~Sun, Semi-supervised
  conditional gans, arXiv preprint arXiv:1708.05789 (2017).

\bibitem{miyato2018cgans}
T.~Miyato, M.~Koyama, cgans with projection discriminator, arXiv preprint
  arXiv:1802.05637 (2018).

\bibitem{liang2018generative}
K.~J. Liang, C.~Li, G.~Wang, L.~Carin, Generative adversarial network training
  is a continual learning problem, arXiv preprint arXiv:1811.11083 (2018).

\bibitem{abrahams1965accuracy}
S.~Abrahams, J.~Bernstein, Accuracy of an automatic diffractometer.
  {M}easurement of the sodium chloride structure factors, Acta
  Crystallographica 18~(5) (1965) 926--932.

\bibitem{yan2013labeb}
X.~Yan, S.~Luo, Z.~Lin, Y.~Yue, X.~Wang, L.~Liu, C.~Chen, {LaBeB}$_3${O}$_7$: A
  new phase-matchable nonlinear optical crystal exclusively containing the
  tetrahedral {XO}$_4$ ({X}={B} and {Be}) anionic groups, Journal of Materials
  Chemistry C 1~(22) (2013) 3616--3622.

\bibitem{idemoto2004relation}
Y.~Idemoto, H.~Yoshikoshi, N.~Koura, K.~Takeuchi, J.~W. Richardson, C.~K.
  Loong, Relation between the crystal structure, physical properties and
  ferroelectric properties of {PbZr}$_x${Ti}$_{1-x}${O}$_3$ (x=0.40, 0.45,
  0.53) ferroelectric material by heat treatment, Journal of the Ceramic
  Society of Japan 112~(1301) (2004) 40--45.

\bibitem{chen2018pzt}
Y.~Chen, X.~Bao, C.-M. Wong, J.~Cheng, H.~Wu, H.~Song, X.~Ji, S.~Wu, {PZT}
  ceramics fabricated based on stereolithography for an ultrasound transducer
  array application, Ceramics International 44~(18) (2018) 22725--22730.

\bibitem{hikam2018pyroelectric}
M.~Hikam, I.~Irzaman, H.~DarmasetiawanArifin, P.~Arifin, M.~Budiman,
  M.~Barmawi, Pyroelectric properties of lead zirconium titanate
  ({PbZr}$_{0.525}${Ti}$_{0.475}${O}$_3$) metal-ferroelectric-metal capacitor
  and its application for {IR} sensor, Jurnal Sains Materi Indonesia 6~(3)
  (2018) 23--27.

\bibitem{burst2016cdte}
J.~M. Burst, J.~N. Duenow, D.~S. Albin, E.~Colegrove, M.~O. Reese, J.~A.
  Aguiar, C.-S. Jiang, M.~Patel, M.~M. Al-Jassim, D.~Kuciauskas, et~al., {CdTe}
  solar cells with open-circuit voltage breaking the 1 {V} barrier, Nature
  Energy 1~(3) (2016) 1--8.

\bibitem{weill2019adanet}
C.~Weill, J.~Gonzalvo, V.~Kuznetsov, S.~Yang, S.~Yak, H.~Mazzawi, E.~Hotaj,
  G.~Jerfel, V.~Macko, B.~Adlam, M.~Mohri, C.~Cortes, {AdaNet}: A scalable and
  flexible framework for automatically learning ensembles (2019).
\newblock \href {http://arxiv.org/abs/1905.00080} {\path{arXiv:1905.00080}}.

\bibitem{molino2019ludwig}
P.~Molino, Y.~Dudin, S.~S. Miryala, Ludwig: a type-based declarative deep
  learning toolbox, arXiv preprint arXiv:1909.07930 (2019).

\bibitem{he2019automl}
X.~He, K.~Zhao, X.~Chu, Auto{ML}: A survey of the state-of-the-art, arXiv
  preprint arXiv:1908.00709 (2019).

\bibitem{one-shot_repo}
J.~M. Ede, One shot exit wavefunction reconstruction, online:
  \url{https://github.com/Jeffrey-Ede/one-shot} (2019).

\bibitem{abadi2016tensorflow}
M.~Abadi, P.~Barham, J.~Chen, Z.~Chen, A.~Davis, J.~Dean, M.~Devin,
  S.~Ghemawat, G.~Irving, M.~Isard, et~al., Tensorflow: A system for
  large-scale machine learning., in: OSDI, Vol.~16, 2016, pp. 265--283.

\bibitem{ede2019improving}
J.~M. Ede, R.~Beanland, Improving electron micrograph signal-to-noise with an
  atrous convolutional encoder-decoder, Ultramicroscopy 202 (2019) 18--25.

\bibitem{wang2019characterization}
C.~Wang, C.~Luo, X.~Wu, Characterization and dynamic manipulation of graphene
  by in situ transmission electron microscopy at atomic scale, Handbook of
  Graphene: Physics, Chemistry, and Biology (2019) 291.

\bibitem{mendes2019electron}
R.~G. Mendes, J.~Pang, A.~Bachmatiuk, H.~Q. Ta, L.~Zhao, T.~Gemming, L.~Fu,
  Z.~Liu, M.~H. R{\"u}mmelii, Electron-driven in situ transmission electron
  microscopy of 2{D} transition metal dichalcogenides and their 2{D}
  heterostructures, ACS nano 13~(2) (2019) 978--995.

\bibitem{zhang2018atomic}
D.~Zhang, Y.~Zhu, L.~Liu, X.~Ying, C.-E. Hsiung, R.~Sougrat, K.~Li, Y.~Han,
  Atomic-resolution transmission electron microscopy of electron
  beam--sensitive crystalline materials, Science 359~(6376) (2018) 675--679.

\bibitem{lakshman2019application}
M.~Lakshman, Application of conventional electron microscopy in aquatic animal
  disease diagnosis: A review, Journal of Entomology and Zoology Studies 7
  (2019) 470--475.

\bibitem{ogawa2019transmission}
Y.~Ogawa, J.-L. Putaux, Transmission electron microscopy of cellulose. {P}art
  2: technical and practical aspects, Cellulose 26~(1) (2019) 17--34.

\bibitem{perez2017learning}
E.~Perez, H.~de~Vries, F.~Strub, V.~Dumoulin, A.~Courville, Learning visual
  reasoning without strong priors, arXiv preprint arXiv:1707.03017 (2017).

\bibitem{gms_webpage}
Gatan, Gatan microscopy suite, online:
  \url{www.gatan.com/products/tem-analysis/gatan-microscopy-suite-software}
  (2019).

\bibitem{adobe1992tiff}
A.~D. Association, et~al., {TIFF} revision 6.0, online:
  \url{www.adobe.io/content/dam/udp/en/open/standards/tiff/TIFF6.pdf} (1992).

\bibitem{tshitoyan2019unsupervised}
V.~Tshitoyan, J.~Dagdelen, L.~Weston, A.~Dunn, Z.~Rong, O.~Kononova, K.~A.
  Persson, G.~Ceder, A.~Jain, Unsupervised word embeddings capture latent
  knowledge from materials science literature, Nature 571~(7763) (2019) 95--98.

\bibitem{beach2018qucumber}
M.~J. Beach, I.~De~Vlugt, A.~Golubeva, P.~Huembeli, B.~Kulchytskyy, X.~Luo,
  R.~G. Melko, E.~Merali, G.~Torlai, Qu{C}umber: {W}avefunction reconstruction
  with neural networks, arXiv preprint arXiv:1812.09329 (2018).

\bibitem{carleo2019netket}
G.~Carleo, K.~Choo, D.~Hofmann, J.~E. Smith, T.~Westerhout, F.~Alet, E.~J.
  Davis, S.~Efthymiou, I.~Glasser, S.-H. Lin, et~al., Net{K}et: {A} machine
  learning toolkit for many-body quantum systems, arXiv preprint
  arXiv:1904.00031 (2019).

\bibitem{salimans2016improved}
T.~Salimans, I.~Goodfellow, W.~Zaremba, V.~Cheung, A.~Radford, X.~Chen,
  Improved techniques for training {GAN}s, in: Advances in neural information
  processing systems, 2016, pp. 2234--2242.

\bibitem{brock2018large}
A.~Brock, J.~Donahue, K.~Simonyan, Large scale {GAN} training for high fidelity
  natural image synthesis, arXiv preprint arXiv:1809.11096 (2018).

\bibitem{FTSR}
H.~Research, {FTSR} software, online:
  www.hremresearch.com/Eng/plugin/FTSREng.html (2019).

\bibitem{MATLAB:2018}
MATLAB, version 9.5 ({R}2018b), The MathWorks Inc., Natick, Massachusetts,
  2018.

\end{thebibliography}

\section{ Acknowledgements }

Thanks go to Christoph T. Koch for software used to collect experimental focal series, to David Walker for suggesting materials in fig.~\ref{fig:unseen_materials}, and to Jessica Marshall for feedback on fig.~\ref{fig:unseen_materials}.

Funding: J.M.E. acknowledges EPSRC grant EP/N035437/1 and EPSRC Studentship 1917382 for financial support, R.B. acknowledge EPSRC grant EP/N035437/1 for financial support, J.J.P.P. acknowledges EPSRC grant EP/P031544/1 for financial support, and J.S. acknowledges EPSRC grant EP/R019428/1 for financial support. 


\setcounter{section}{0}
\renewcommand{\thesection}{Appendix \Alph{section}}

\section{Sharded Deep Holography}\label{appendix:sharded}

Collecting experimental CTEM holograms with a biprism or focal series reconstruction is expensive: Measuring a large number of representative holograms is time-intensive, and requires skilled electron microscopists to align and operate microscopes. In this context, we propose a new method to reconstruct holograms by extracting information from a large image database with deep learning. It is based on the idea that individual images are fragments of aberration series sampled from an aberration series distribution. To be clear, this section summarizes an idea and is intended to be a starting point for future work.

Let $\psi_\text{exit} \sim \Psi_\text{exit}$ denote an unknown exit wavefunction, $\psi_\text{exit}$, sampled from a distribution, $\Psi_\text{exit}$, $c \sim  C$ denote an unknown contrast transfer function (CTF), $c = \psi_\text{pert}(\textbf{q})/\psi_\text{dif}(\textbf{q})$, sampled from a distribution, $C$, and $m \sim M$ denote metadata, $m$, sampled from a distribution, $M$, that restricts $\Psi_\text{exit}$. The image wave is
\begin{align}
\psi_\text{img} = \text{FT}^{-1}(c \text{FT}(\psi_\text{exit})).
\end{align}
We propose introducing a faux CTF, $c' \sim C'$, to train a cycle-consistent generator, $G$, and discriminator, $D$, to predict the exit wave, 
\begin{equation}
\hat{\psi}_\text{exit} = G(|\psi_\text{img}|, m).
\end{equation}
The faux CTF can be used to generate an image wavefunction
\begin{equation}
\hat{\psi}_\text{img}' = \text{FT}^{-1}(c' \text{FT}(\hat{\psi}_\text{exit}) ).
\end{equation}
If the faux distribution is realistic, $D$ can be trained to discriminate between $|\hat{\psi}_\text{img}'|$ and $|\psi_\text{img}|$. For example, by minimizing the expected value of
\begin{equation}
L_D = D(|\psi_\text{img}|, m)^2 + (D(|\hat{\psi}_\text{img}'|, m') - 1)^2,
\end{equation}
where $m' \neq m$ if metadata describes different CTFs. A cycle-consistent adversarial generator can then be trained to minimize the expected value of
\begin{equation}\label{eqn:exp_holo_L_G}
\begin{split}
L_G = \ &D(|\hat{\psi}_\text{img}'|, m)^2 \ + \\
&\lambda ||G(|\psi_\text{img}|, m) - G(|\hat{\psi}_\text{img}'|, m')||^2_2,
\end{split}
\end{equation}
where $\lambda$ weights the contribution of the adversarial and cycle-consistency losses. The adversarial loss trains the generator to produce realistic wavefunctions, whereas the cycle-consistency loss trains the generator to learn unique solutions.

Alternatively, CTFs could be preserved by mapping
\begin{equation}
G(|\hat{\psi}_\text{img}'|, m) \rightarrow \text{FT}^{-1}(\text{FT}(G(|\hat{\psi}_\text{img}'|, m))/c'),
\end{equation}
when calculating the L2 norm in eqn.~\ref{eqn:exp_holo_L_G}. If CTFs are preserved by this mapping, $c'$ is a relative; rather than absolute, CTF and $cc'$ is the CTF of $\hat{\psi}_\text{img}'$. 

Two of our experimental datasets containing 17267 TEM and 16227 STEM images are available with our new wavefunction datasets\cite{warwickem!}. However, the images are unlabelled to anonymise contributors; limiting metadata available to restrict a distribution of wavefunctions.

\section{Experimental Focal Series}\label{appendix:experimental}

As a potential starting point for experimental one-shot exit wavefunction reconstruction, we have made 1000 focal series publicly available\cite{warwickem!}. We have also made simple focal series reconstruction code available at \cite{one-shot_repo}. Alternatively, refined focal and tilt series reconstruction (FTSR) software is commercially available\cite{FTSR}. Each series consists of 14 32-bit 512$\times$512 TIFFs, area downsampled from 4096$\times$4096 with MATLAB\cite{MATLAB:2018} and default antialiasing. All series were created with a common, quadratically increasing\cite{haigh2013recording} defocus series. However, spatial scales vary and must be fitted as part of reconstruction.

\end{document}